\documentclass[]{emulateapj}
\slugcomment{Astrophysical Journal Letters, accepted}
\shorttitle{On the planet-metallicity correlation}
\shortauthors{Haywood}

\begin{document}

\title{On the correlation between metallicity and the presence of giant planets}
\author{Misha Haywood}

\affil{GEPI, Observatoire de  Paris, CNRS, Universit\'e Paris Diderot; 92190 Meudon,France
}
\email{Misha.Haywood@obspm.fr}

\begin{abstract}
The correlation between stellar  metallicity and the presence of giant
planets is well established. It  has been tentatively explained by the
possible  increase of  planet formation  probability in  stellar disks
with enhanced  amount of  metals.  However, there  are two  caveats to
this  explanation.   First, giant  stars  with  planets  do not show  a
metallicity distribution  skewed towards metal-rich  objects, as found
for dwarfs.  Second, the correlation  with metallicity is not valid at
intermediate metallicities,  for which it can be shown  that giant planets
are preferentially found orbiting thick disk stars.

None of these two peculiarities is explained by the proposed scenarios
of  giant planet  formation.  We  contend  that they  are galactic  in
nature,  and probably  not linked  to the  formation process  of giant
planets.  It is  suggested that the same dynamical  effect, namely the
migration of  stars in  the galactic  disk, is at  the origin  of both
features,  with the important consequence that most  metal-rich stars
hosting giant planets  originate from the inner disk,  a property that
has  been   largely  neglected  until  now.   We   illustrate  that  a
planet-metallicity correlation  similar to the observed  one is easily
obtained  if stars from  the inner  disk have  a higher  percentage of
giant planets  than stars born at  the solar radius,  with no specific
dependence on metallicity.  We propose that the density  of $\rm H_2$
in the inner  galactic disk (the molecular ring) could  play a role in
setting the high percentage of  giant planets that originate from this
region.
\end{abstract}


\section{Introduction}

Metal-rich stars ([Fe/H]$>$+0.25 dex) found in the solar neighborhood, including 
giant planet hosts, are objects that have migrated from the inner disk (i.e inside the solar galactocentric radius) by the 
effect of radial mixing (Sellwood \& Binney 2002; Haywood 2008b). This is 
an intriguing fact, but
since it has been proposed that the prevalence of jovian planets on metal-rich stars could  
be due to the enhanced probability of forming planetesimals in an environment enriched in metals, 
why should one bother about the galactic origin of the host stars?
The first reason is that the correlation between stellar metallicities and the presence of
giant planets  is made up of stars of different galactic origins, 
which could bear some importance on the formation of giant planets themselves.
The second is that the study of the effect of radial mixing permits new insights on two particularities that otherwise do not fit well 
into the scenario of metal-enhanced planet formation, namely the 'normal' metallicity distribution 
of giants stars with planets (Pasquini et al. (2007), Takeda et al. (2008)), and the fact that, at intermediate metallicities, giant planets seems to 
favor thick disk stars rather than thin disk objects (Haywood, 2008a).

Both arguments are the subject of this Letter,
which is laid out as follows: in $\S$2 we briefly discuss the evidences of radial mixing and
show how metallicity  depends on the galactic orbital parameters of planet host stars.
In \S3, we show that (1) the difference in the metallicity distribution of planet host giants and dwarfs
and (2) the difference in the number of planet hosts between the thin and thick disks can be both explained as a consequence of radial
mixing. In \S4, we discuss the implication of these results, and demonstrate that if the 
percentage of giant planets among stars is dependent upon the galactocentric distance, 
a correlation between stellar metallicity and the presence of planet is a natural 
outcome of radial migration. Finally, we discuss what galactic property, dependent on galactocentric distance, 
could explain these new results.

\section{Radial mixing in the galactic disk}

\begin{figure*}
\plotone{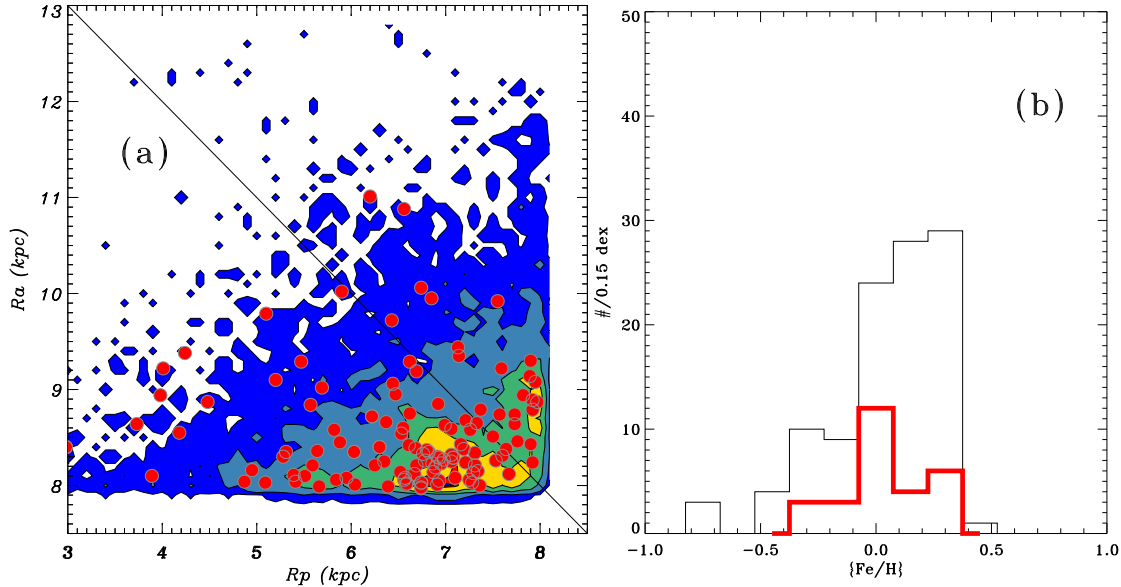}
\caption{On the left, a density plot of the apocenter $vs$ pericenter of stars in the GCS catalogue.
The diagonal line represents a mean orbital radius of 8 kpc. Stars with planets and orbital parameters in the GCS 
catalogue are plotted as large dots.  On the right, the histograms of host planet stars that have R$_{mean}$=(R$_p$+R$_a$)/2.$<$8 kpc
(thin line) or   R$_{mean}>$8 kpc (thick red line).
\label{fig:amr}}
\end{figure*}

\subsection{Evidences of radial mixing}

The suggestion that stars migrate in the galactic disk has been around since at least the seventies.
A few dynamical processes have been claimed responsible for this phenomenon: random scattering (Wielen 1977), 
'churning' by spirals (Sellwood \& Binney  2002) or perturbations by an orbiting satellite (Quillen et al. 2009), 
but specific observational evidence for any such processes remain elusive. 
The only direct evidence so far that such processes may be active is the differentiation  
encrypted in the orbital parameters of solar neighborhood stars considered as a function of
metallicity. An example is given in Fig.~4 of Haywood (2008b), which shows the clear difference in the 
distribution of apo and pericenters  for metal-poor and metal-rich stars. 
Other indirect evidences exist. One is the increasing metallicity dispersion with age 
(Haywood 2006, 2008b), which testifies that the older the star, the greater 
the distance over which it may have migrated, and therefore come from a region with significantly different
mean stellar abundance. The second is that the 'terminal' metallicity reached by the local chemical evolution 
is about 0.2 dex, not 0.4 or 0.6 dex.
This is evidenced by the fact that
there are no young stars above [Fe/H]$\approx$0.2 dex in the solar neighborhood. The solar radius has simply not 
reached this state of chemical evolution, and super metal-rich objects must have formed elsewhere, the inner
disk being the most probable site.

\subsection{The impact on the planet host star population}

The galactic aspects of the bias towards metal-rich stars among planet hosts have been seldom investigated.
In particular, the evidence that metal-rich stars of the solar vicinity must 
have come from the inner disk has been barely discussed. The exceptions are Ecuvillon et al. (2007)
and Haywood (2008a). 
Several studies investigated whether planet host stars have properties 
that could differ from those of common field stars, but it 
has been so far rather unfruitful (see Udry \& Santos 2007). 
Robinson et al. (2006) found host stars
to be overabundant in Si and Ni, but this has not been confirmed so far 
(see Gonzalez \& Laws 2007). The only clear evidence of a difference has been reported
in  Haywood (2008a), where it was shown that at intermediate metallicities ([Fe/H]$<$-0.3),
most stars known to harbor giant planets belong to the thick disk rather than to the thin disk (in the ratio 10/2). 
We come back to this point in the next section. In any case, there is no reason to suspect that the metal-rich
host stars are not  being affected by radial mixing and that their origin would be different from other stars 
of the same metallicities.

Is there direct evidence of radial mixing among planet host stars? Fig. 1a shows the pericenter-apocenter
distribution for stars in the GCS (Geneva-Copenhagen Survey) catalogue (Nordstr{\"o}m et al. 2004) as a density-surface plot. 
The analysis of metallicity distribution of the stars in this plot shows that metal-poor thin disk populates
preferentially the upper right part of the diagram (R$_{mean}>$8 kpc), or outer orbits, while the metal-rich 
stars occupy mainly inner orbits (see Haywood 2008b, Fig. 6). This is interpreted as an effect of radial mixing. 
The metal-poor stars show a specific kinematic signature, having a component in the direction of rotation
significantly larger than the LSR (Haywood 2008b). A similar signature have been obtained by Schoenrich \& Binney (2009) in 
modeling the effect of 'churning', which basically allows stars to swap between circular orbits of different angular 
momentum. Their Fig. 3 shows that within a few Gyr, the mean orbital radius of a star can change by several kpc, 
both inward or outward. So we do expect that stars coming from inside the solar circle (respectively outside)  
to populate inner (outer) orbits. More details on the observational signatures and other consequences are given in Haywood (2008b). 
Sellwood \& Binney (2002) describe the effect of 'churning' while  Schoenrich \& Binney (2009) discuss the chemical evolution 
aspects. 
Red symbols on Fig.1a represent known host-planet stars for which orbital parameters from the GCS catalogue are available. 
The asymmetry in the distribution of orbital parameters is clearly
apparent in the stars with detected planets, with the overwhelming number of objects (79\%) having R$_{\rm mean}=(\rm R_p+R_a)/2<$ 8 kpc.
The histograms on the right illustrate the metallicity distribution for the two groups of stars.
It clearly shows that stars with planets are subject to a differentiation, the group with  R$_{\rm mean}<$8 kpc having
a metallicity distribution similar to dwarfs hosting planets in general, while the group at R$_{\rm mean}>$8 kpc 
has a metallicity centered on [Fe/H]=0. 
It shows that planet hosts follow the general behavior of the metal rich population and that the specific high metallicities can reasonably be attributed 
to a common origin in the inner disk.

\section{Two pitfalls}

After the discovery that the presence of Jovian planets is more frequent around metal-rich 
stars, several studies have explored how the formation of giant planets could be
favored in a circumstellar disk enhanced in metals (Ida \& Lin 2004, Mordasini et al. 2008).  
The results of the previous section now leads to the following preliminary 
question: how do we know that the higher percentage
of giant planets detected on metal-rich stars is due to their metallicity and not to some other factor 
also linked with their origin in the inner disk ? 
The question is relevant, because any measurable property of inner disk stars other than metallicity would be 
correlated with the presence of
planet. The obvious a priori response is that metallicity is a measurable 
parameter, and  intrinsic to the star. But there could be others however, which, although not measurable on the stars, 
could be no less important, such as, for example, the surface density of molecular hydrogen in the inner galactic disk regions.
We now show
that there are two cases where the planet-metallicity correlation is not verified, for which radial mixing provides a simple 
explanation, suggesting that metallicity may not be the relevant parameter. We will show in the next section that a bona-fide planet-metallicity correlation can be obtained
in the context of radial mixing with no metallicity dependence of any kind.

\subsection{The difference in giant-dwarf metallicity distributions}

The first case where the planet-metallicity correlation breaks down is the metallicity distribution of giant
host stars. While the planet-host dwarf metallicity distribution is known to be skewed towards metal-rich
objects, giant hosts are known to have a metallicity which is more like the field
stars distribution, see in particular Takeda et al. (2008) and Pasquini et al. (2007).

\begin{figure}
\plotone{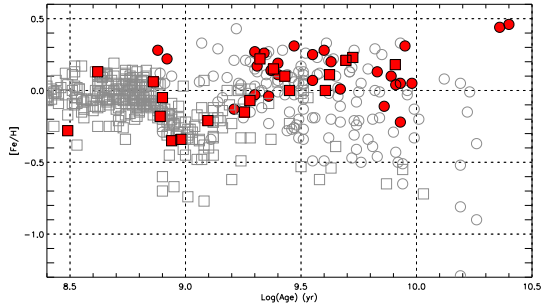}
\caption{The age-metallicity relation for giants (squares) and dwarfs (circles) from Takeda (2007) 
and Takeda et al. (2008), complemented  by 'massives' (M$>$1.4 M$_{\odot}$)  objects from the exoplanet database.
Host planet dwarfs are shown as red large dots, host planet giants as red filled squares. 
The mean metallicity of 14 planet host stars with ages $<$ 2 Gyr is -0.04 dex.
\label{fig:amr}}
\end{figure}

Figure \ref{fig:amr} shows the age-metallicity relation for field dwarfs and giants as derived
by Takeda (2007), and Takeda et al. (2008). The progressive enlargement of the metallicity with age is well 
in accord with the effect of radial mixing, as noted in Haywood (2006, 2008b).
The metallicity dispersion is smaller for giants: this is expected given their age
distribution, Fig. \ref{fig:amr} showing that most giants have ages smaller than 1 Gyr.
Since radial mixing is a secular process, its effect increases with time: the contamination
by stars from the inner and outer disk is proportional to age. Because the sample of giants contains 
mostly young stars, it is little polluted by old, metal-rich, wanderers of the inner disk.
The squares on Fig. \ref{fig:amr} show the planet host giant stars from Takeda et al. (2008), completed with the 
few other 'massive' (M$>$1.4 M$_{\odot}$) objects available from the exoplanet database\footnote{J. Schneider, http://exoplanet.eu}.
The figure illustrates that these objects, when older than about 2 Gyr, are mostly metal-rich ([Fe/H]$>$0.0 dex),
while the 14 stars younger than 2 Gyr have a mean metallicity of -0.04 dex.
In the sample of Takeda et al., 7 host giants (out of 10) are younger than 1 Gyr, and all are younger than 3 Gyr. 
The explanation for the difference between the dwarf and giant distributions comes out as a natural 
galactic effect: the giant sample contains a limited bias towards metal-rich objects because it is much younger
than the dwarf sample, and then much less contaminated by radial mixing.  

Pasquini et al. (2007) have suggested that the mass of the convective envelope could 
play a role. If the excess of metals is due to pollution at the surface of stars, it could be diluted
when the dwarf becomes a giant.
We propose instead that the excess of metals is intrinsic to the star, and that the age is the determining
factor, producing a selective effect on the origin of the stars.

\begin{figure}
\plotone{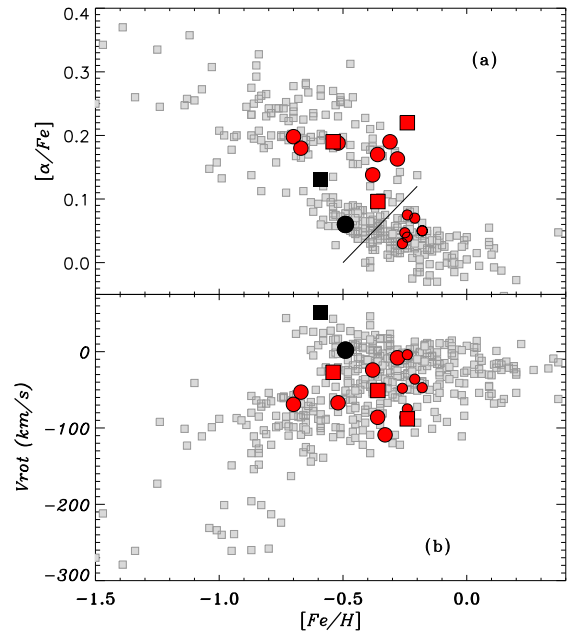}
\caption{(a) Stars with giant planets at [Fe/H]$<$-0.2 dex for which $\alpha$ 
element abundance is available (the mean of Mg, Si, Ca, Ti, or the last three for some giants). 
Grey squares are field dwarfs from Reddy et al. (2003, 2006) and Bensby et al. (2005).
Large red dots or square symbols are host planet dwarfs or giants in the thick disk regime 
and transition zone between the thick and thin disks. 
The black dot and square represent the only dwarf (HD 171028) and giant hosting jupiters
clearly in the metal-poor thin disk regime ([Fe/H]=-0.49, -0.59 dex)
and V$_{\rm rot}$=(+2, 51.5) km.s$^{-1}$.
The smaller dots below the line are dwarfs clearly in the thin disk regime, with their lag 
in V$_{rot}$ suggesting they are not from the outer disk. 
Plot (b) shows the velocity component in the 
direction of rotation as a function of metallicity for stars in plot (a). 
The field stars that make up the branch towards V$_{rot}>$0 and low metallicities ([Fe/H]$<$-0.3 dex)
are the metal-poor objects with a probable origin in the outer disk (at [Fe/H]$<$-0.3 dex and [$\alpha$/Fe]$<$0.1 dex in plot (a)).   
\label{fig:fehvrot}}
\end{figure}

\subsection{The difference between the thin and thick disks}

At intermediate metallicities (-0.7$<$[Fe/H]$<$-0.3 dex), stellar populations in the solar vicinity can be divided
into two groups: the thin and the thick disks, which differentiate both by their $\alpha$-elements
content and their asymmetric drift. At these metallicities, the thin disk is solar in $\alpha$-elements, 
but rotates faster than the LSR, while the thick disk is enriched in $\alpha$-elements ([$\alpha$/Fe]$>$0.1 dex) but lags the 
LSR.
While the local metal-rich stars may be attributed to migration 
from the inner disk, the metal-poor end can be attributed to
stars that came from the outer disk (see Haywood 2008b).
It has been shown in Haywood (2008a) that in this metallicity interval, giant planets are found
preferentially on thick disk stars. This is illustrated in Fig. \ref{fig:fehvrot}, where 10 stars with 
giant planets are compatible with being either thick disk or transition objects between the thin and 
the thick disks. Only 1 dwarf, HD 171028, and 1 giant, HD 170693, are compatible with being a member of 
the metal-poor thin disk with an origin in the outer disk. 
As commented in Haywood (2008a), this is significant, because the number of thin disk objects at these metallicities
is expected to be higher or equal to the number of thick disk stars. 

On  Fig. \ref{fig:fehvrot}a, 6 objects having [Fe/H]$<$-0.2 dex are thin disk objects (smaller symbols 
below the line on plot (a)). The rotation lag and $\alpha$-element content of these stars (plot b)
support the view that they are bona-fide solar radius objects, with no specific indication that they would come from 
the outer disk.
The search of new giant planet hosts in this metallicity range with no bias in favor or against either the thin disk and the 
thick disk is highly desirable to confirm this trend, but we think the difference between the two groups is significant.

Finally, it should be noted that the galactocentric radii of origin of thick disk stars (those with [Fe/H]$<$-0.2 dex 
and [$\alpha$/Fe]$>$0.15 dex on Fig.  \ref{fig:fehvrot}) is not clear.  According to  Schoenrich \& Binney (2009), we should 
expect the most metal-rich (at [Fe/H]$>$-0.4,-0.5 dex) and $\alpha$-elements enhanced thick disk objects to come from the 
inner disk. This could be the case in particular for HIP 3497, HIP 26381, HIP 58952, 
HIP 62534, which all have metallicities above [Fe/H]=-0.5, and relatively high level of $\alpha$ abundance.
This is an interesting possibility since in this eventuality even thick disk objects could originate from the inner disk.

If metallicity was the determining factor for the presence of giant planet, we should not expect 
a difference between the number of planet host stars of the thin and thick disks. Since the metal-poor 
thin disks objects are expected to come from the outer disk, it is again suggested that the distance to the galactic center plays a role.

\section{Discussion}

We are now facing the following picture: stars that come from 
the inner disk are noticeably rich in giant planets, while stars that come from the outer disk seems
to be less favored in this respect. 
This new information changes considerably how we envisage the correlation between
metallicity and the presence of giant planets. 
For the surprising point here is not the fact that most host-planet stars are metal-rich, since they come from a region 
where most stars are metal-rich, but the very fact that most would come from the inner disk.
We are led to conclude that the distance to the galactic center must somehow play a role in 
setting the percentage of giant planets, with two new questions : (1) what of the correlation between stellar 
metallicity and planet, and (2) what is the parameter linked to the galactocentric distance which could
influence the percentage of giant planets ?

\begin{figure}
\plotone{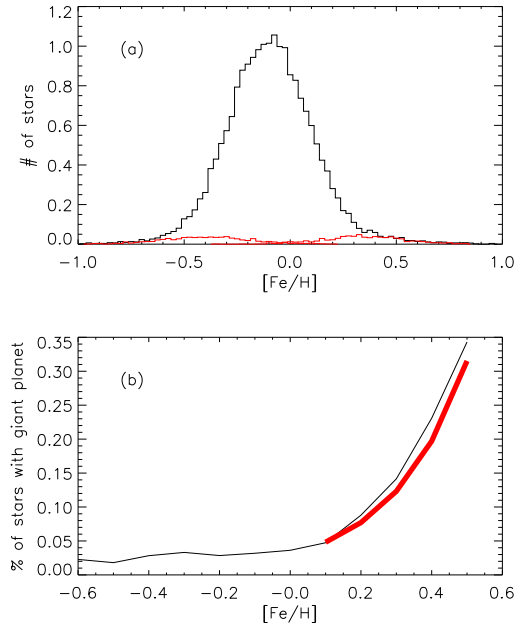}
\caption{(a) Simulated 'local' metallicity distribution.
In red, the contributions of the metal-rich and metal-poor components assumed
to have come to the solar neighborhood by radial migration. See section 4 
for details.
(b) The percentage obtained assuming the metallicity distribution and intrinsic 
giant planet proportion of 0\% in the metal-poor component, 5\% locally,  
25\% in the metal rich component. The thick line is the 
percentage of planet host $vs$ stellar metallicity according to the fit 
given by Udry \& Santos (2007).
\label{fig:fehdist}
}
\end{figure}

(1) In this new scheme, the well-admitted correlation
between metals and the presence of giant planets could be the mere reflection of the galactic origin of the stars, 
but does not necessarily implies an effect of the metallicity on the formation of giant planets. 
To illustrate this prediction, we make the following simple estimate. 
We adopt a 'local'  (e.g for stars born at the solar galactocentric radius)  gaussian metallicity distribution 
centered on [Fe/H]=-0.1 dex, with intrinsic dispersion 0.1 dex. 
We assume that about 4\% (Grenon 1989) of the stars at the solar radius come from the inner disk, sampling a
metallicity distribution centered on [Fe/H]=+0.35, with dispersion +0.2 dex (a higher dispersion 
takes into account, in a simplified way, the fact that stars come from different inner radii, and therefore from regions
where the mean metallicity is not strictly 0.35 dex). Given a metallicity 
gradient of about 0.07 to 0.1 dex.kpc$^{-1}$, which is about what is measured (Maciel \& Costa 2008) a mean metallicity [Fe/H]=+0.35 dex can be expected towards 
the galactic center at about 3 to 5 kpc from the Sun. 
We assume that the percentage of host-planet stars in the inner disk is 25\% (as measured on the most 
metal-rich objects of the solar neighborhood) and {\it independent} of metallicity.
We also assume a 4\% of metal-poor stars centered on [Fe/H]=-0.4 dex, with a dispersion in metallicity of +0.2 dex, 
with no giant planets. 
Finally, an error on measured metallicities is simulated with a random gaussian with 0.15 dex dispersion.
The metallicity distribution generated with these parameters is given in Fig.\ref{fig:fehdist} (a). 
On plot (b), we show the proportion of stars with giant planets obtained with our assumptions. 
The thick line is the fit made by Udry \& Santos (2007) on the observational distributions (3.01$\times$10$^{2.04[Fe/H]}$).
As can be seen, a good correlation between the presence of a giant planet and the metallicity of
the stars is obtained, providing an honest fit to the observed rate.

(2) Some factor linked to galactocentric distance,
but not metallicity, must play a role in setting the percentage of giant planet. 
A candidate could be the density of dust in the inner disk, because dust is thought to favor the formation of
planetesimals. However, there is, as yet, no evidence for 
a difference between the distributions of dust and metals in our Galaxy, so that we do not expect dust 
to lead to different patterns than metallicity.
A better candidate is molecular hydrogen. It is foremost a fundamental ingredient for the formation of giant 
planets, being the principal constituent of stellar disks and Jupiters. 
Its main structure in the Galaxy, the molecular ring, is thought to contain 70\% of  
H$_2$ gas inside the solar circle (Clemens et al. 1988, Jackson et al. 2006), thereby providing a huge reservoir
for star (H$_2$ is known to be directly linked to star formation (Kennicutt 2008)) and planet formation. 
The most interesting aspect however, is the fact the molecular ring reaches a maximum density at 3-5 kpc 
from the sun, corresponding to the distance 
where stars with metallicity in the range (+0.3,+0.5) dex are expected to be formed preferentially.
Interestingly, the mean surface or volume density of H$_2$ at a galactocentric distance of 4 to 5 kpc 
is 2 to 5 times its local value (Nakanishi \& Sofue 2006), in proportion with the rate of giant planets on metal-rich 
(25\% (roughly $\pm$ 10\%)) and solar metallicity stars (4\%).

A final indication may comes to support our views. Stars hosting only Neptunian and/or super-Earth should 
be less prone to having an 
origin in the inner disk, if they can form in environment less dense in H$_2$. 
Which means that we should not expect a predominance of metal-rich stars among Neptunian/super-Earth host stars.
Among the 12 objects on which Neptunian or super-earth planets have been discovered, 7 with
no Jovian planets have metallicities -0.28, -0.33, -0.31, -0.31, -0.05, -0.1, and -0.15 (GJ~674, Gl~581, HD~4308, HD~40307, HD~69830, HD 285968, HD 7924) according to the exoplanet
database. The 5 stars also harboring Jovian planets (HD 75732, Gl 876, HD 47186, HD 160691, HD 181433) have metallicities 
+0.29, -0.12, +0.23, +0.28 and +0.33 dex, amply confirming the possibility that the first group of stars could be genuine 
solar radius objects, and the second  wanderers from inside the Galaxy.

\acknowledgements

I thank the referee for a prompt report and helpful comments 
and A.-L. Melchior for her suggestions and comments.


\begin{thebibliography}{}
\bibitem[Bensby et al.(2005)]{ben05} Bensby T., Feltzing S., Lundstr{\"o}m I., Ilyin I., 2005, A\&A, 433, 185 
\bibitem[Clemens, Sanders, \& Scoville(1988)]{1988ApJ...327..139C} Clemens D.~P., Sanders D.~B., Scoville N.~Z., 1988, ApJ, 327, 139 
\bibitem[Ecuvillon et al.(2007)]{2007A&A...461..171E} Ecuvillon A., Israelian G., Pont F., Santos N.~C., Mayor M., 2007, A\&A, 461, 171 
\bibitem[Gilli et al.(2006)]{2006A&A...449..723G} Gilli G., Israelian G., Ecuvillon A., Santos N.~C., Mayor M., 2006, A\&A, 449, 723 
\bibitem[Gonzalez \& Laws(2007)]{2007MNRAS.378.1141G} Gonzalez G., Laws C., 2007, MNRAS, 378, 1141 
\bibitem[Grenon(1989)]{1989Ap&SS.156...29G} Grenon M., 1989, Ap\&SS, 156, 29 
\bibitem[Haywood(2006)]{2006MNRAS.371.1760H} Haywood M., 2006, MNRAS, 371, 1760 
\bibitem[Haywood(2008)]{2008A&A...482..673H} Haywood M., 2008a, A\&A, 482, 673 
\bibitem[Haywood(2008)]{2008MNRAS.388.1175H} 
Haywood M., 2008b, MNRAS, 388, 1175 
\bibitem[Ida \& Lin(2004)]{2004ApJ...616..567I} Ida S., Lin D.~N.~C., 2004, ApJ, 616, 567 
\bibitem[Jackson et al.(2006)]{2006ApJS..163..145J} Jackson J.~M., et al., 2006, ApJS, 163, 145 
\bibitem[Kennicutt(2008)]{2008ASPC..390..149K} 
Kennicutt R.~C., Jr., 2008, ASPC, 390, 149 
\bibitem[Maciel \& Costa(2008)]{2008arXiv0806.3443M} Maciel W.~J., Costa R.~D.~D., 2008, arXiv, arXiv:0806.3443 
\bibitem[Mordasini et al.(2008)]{2008ASPC..398..235M} Mordasini C., Alibert Y., Benz W., Naef D., 2008, ASPC, 398, 235 
\bibitem[Nakanishi \& Sofue(2006)]{2006PASJ...58..847N} Nakanishi H., Sofue Y., 2006, PASJ, 58, 847 
\bibitem[Nordstr{\"o}m et al.(2004)]{2004A&A...418..989N} Nordstr{\"o}m B., et al., 2004, A\&A, 418, 989 
\bibitem[Pasquini et al.(2007)]{2007A&A...473..979P} Pasquini L., D{\"o}llinger M.~P., Weiss A., Girardi L., Chavero C., Hatzes A.~P., da Silva L., Setiawan J., 2007, A\&A, 473, 979
\bibitem[Quillen et al.(2009)]{2009arXiv0903.1851Q} Quillen A.~C., Minchev I., Bland-Hawthorn 
J., Haywood M., 2009, arXiv, arXiv:0903.1851 
\bibitem[Reddy et al.(2003)]{2003MNRAS.340..304R} 
Reddy B.~E., Tomkin J., Lambert D.~L., Allende Prieto C., 2003, MNRAS, 340, 304 
\bibitem[Reddy, Lambert, \& Allende Prieto(2006)]{2006MNRAS.367.1329R} Reddy B.~E., Lambert D.~L., Allende Prieto C., 2006, MNRAS, 367, 1329 
\bibitem[Robinson et al.(2006)]{2006ApJ...643..484R} Robinson S.~E., Laughlin G., Bodenheimer 
P., Fischer D., 2006, ApJ, 643, 484 
\bibitem[Schoenrich \& Binney(2008)]{2008arXiv0809.3006S} Schoenrich R., Binney J., 2008, arXiv, arXiv:0809.3006 
\bibitem[{Sellwood \& Binney}{2002}]{2002MNRAS.336..785S} Sellwood J.~A., Binney J.~J., 2002, MNRAS, 336, 785 
\bibitem[Takeda(2007)]{2007PASJ...59..335T} Takeda Y., 2007, PASJ, 59, 335 
\bibitem{2008PASJ...60..781T} Takeda Y., Sato B., Murata D., 2008, PASJ, 60, 781 
\bibitem[Udry \& Santos(2007)]{2007ARA&A..45..397U} Udry S., Santos N.~C., 2007, ARA\&A, 45, 397 
\bibitem[Wielen(1977)]{1977A&A....60..263W} Wielen R., 1977, A\&A, 60, 263 
\end{thebibliography}
\end{document}